\documentclass[%
 reprint,
superscriptaddress,
nofootinbib,amsmath,amssymb,aps,prd
]{revtex4-2}
\usepackage{multirow}
\usepackage{graphicx}
\usepackage{comment}
\usepackage[normalem]{ulem}
\usepackage[dvipsnames]{xcolor}
\definecolor{bluecite}{HTML}{0875b7}
\usepackage[colorlinks=true,urlcolor=Emerald,linkcolor=bluecite,citecolor=bluecite]{hyperref}
\usepackage[nameinlink]{cleveref}

\usepackage{amsmath,amssymb,amsfonts}
\usepackage[T1]{fontenc}
\usepackage[utf8]{inputenc}
\usepackage{bbm}

\newcommand{\be}{\begin{equation}}
	\newcommand{\ee}{\end{equation}}
\newcommand{\bea}{\begin{eqnarray}}
	\newcommand{\eea}{\end{eqnarray}}
\newcommand{\vv}{``}

\usepackage[nolist,nohyperlinks]{acronym}

\begin{document}

\title{On the RG flow of  the Newton and cosmological constant}

\author{Carlo Branchina}
\affiliation{Department of Physics, University of Calabria, and INFN-Cosenza, 
	Arcavacata di Rende, I-87036, Cosenza, Italy}

\author{Vincenzo Branchina}
\affiliation{Department of Physics, University of Catania, and INFN-Catania, 
	Via Santa Sofia 64, I-95123 Catania, Italy}

\author{Filippo Contino}
\affiliation{Scuola Superiore Meridionale, Largo San Marcellino 10, 80138 Napoli, Italy}

\author{Riccardo Gandolfo}
\affiliation{Department of Physics, University of Catania, and INFN-Catania, 
	Via Santa Sofia 64, I-95123 Catania, Italy}

\author{Arcangelo Pernace}
\affiliation{Department of Physics, University of Catania, and INFN-Catania, 
	Via Santa Sofia 64, I-95123 Catania, Italy}
	\affiliation{Centro de Física Teórica e Computacional, Faculdade de Ciências, Universidade de Lisboa}

\begin{abstract}

\noindent
In this note we comment on the RG flow of the Newton and cosmological constants,  also in view of some recent claims \cite{Held:2025vkd} that would rise some doubts on the validity of our recent work  \cite{Branchina:2024xzh,Branchina:2024lai}. Here we show that the arguments and claims of \cite{Held:2025vkd}  are seriously flawed and cannot be trusted.

\end{abstract}

\maketitle

The Einstein-Hilbert (EH)
theory of gravity is   perturbatively non-renormalizable.   One possibility is that it is an effective field theory (EFT), i.e.\,\,a quantum field theory (QFT) valid up to an energy scale  
above which its ultraviolet (UV) completion (maybe string theory) takes over. Another possibility is that it is   non-perturbatively renormalizable, in which case  a (non-trivial)
UV-attractive fixed point should exist:\,\,this is the so-called  asymptotic safety (AS) scenario.
Naturally, even if such a fixed point exists and even if matter fields are included, this does not  
mean that the theory  provides the UV closure of particle physics.
What would be physically relevant in this case  is the way the couplings flow in the neighborhood of this fixed point (think for instance of QCD). For a thorough discussion of these matters, we refer to \cite{Alexandre:1997gj}  (see sections I and II and Fig.1).

In  \cite{Held:2025vkd}, it is claimed  that the existence of the AS fixed point for quantum gravity (QG) is well established. In\,\,\cite{Branchina:2024lai},\,\,we find results that point to  the opposite.
Our work is challenged in \cite{Held:2025vkd}. Below we show that the arguments and claims of \cite{Held:2025vkd}  are flawed. 

QFTs are defined through the path integral 
\begin{equation}\label{pathintegral}
	\int {\rm d}\mu \,e^{-S}\,,
\end{equation}
where ${\rm d}\mu$ is the measure and $S$ the classical action. Both the traditional formulation of QFTs 
and the Wilsonian  renormalization group (RG) approach are based on\,\eqref{pathintegral}.
Obviously, two couples $({\rm d}\mu_1,S_1)$ and $({\rm d}\mu_2,S_2)$ for which  ${\rm d}\mu_1\, e^{-S_1}={\rm d}\mu_2 \,e^{-S_2}$ define the same theory, i.e.\,belong to the same equivalence class. If ${\rm d}\mu_1\, e^{-S_1}\neq{\rm d}\mu_2 \,e^{-S_2}$, in general we have two different theories. 

The authors of   \cite{Held:2025vkd} consider two simple scalar theories in three dimensions, with $S_1=S_2= \int {\rm d}^3 x \,\big[ \frac12 (\partial_\mu \varphi)^2 + \frac{m^2}{2} \varphi^2 + \frac{\lambda}{4!} \varphi^4 \big] $ and ${\rm d}\mu_1\neq{\rm d}\mu_2$, namely ${\rm d}\mu_1= {\rm d}\varphi$ and ${\rm d}\mu_2= {\rm d}\varphi \,e^{-\int {\rm d}^3 x\, \lambda_8 \varphi^8}$. As it is well-known, the theory $({\rm d}\mu_1,S_1)$ is UV-complete (asymptotically free in the present case), while the theory $({\rm d}\mu_2,S_1)$ is not. They claim that for this  UV-incomplete theory \vv no path integral exists'' and that it is  \vv inconsistent''. Furthermore, they claim that the UV-incompleteness of $({\rm d}\mu_2,S_1)$ does
not invalidate the existence of the  asymptotically free
scalar theory $({\rm d}\mu_1,S_1)$.

The first claim is obviously {\it wrong}. It conflicts with the idea of EFT itself. According to Wilson's lesson,  UV-incomplete theories are as welcome as  UV-complete ones. In the former case,  the UV cutoff is the scale above which the theory is replaced by its UV-completion, and the path integral is defined as for UV-complete theories (see above). 
The second claim is  obviously true.

Now comes a delicate point. 
The authors  of \cite{Held:2025vkd} claim that a  definition of a QFT \vv equivalent'' to \eqref{pathintegral}   is provided by the evolution equation for the so-called effective average action $\Gamma_k$ (Eq.\,(5) of\,\,\cite{Held:2025vkd})
\begin{equation}
	k \partial_k \Gamma_k = \frac{1}{2} \text{Tr} \left[ \left( \Gamma_k^{(2)} + R_k \right)^{-1} k \partial_k R_k \right] \, .
	\label{eq:flow}
\end{equation}
This claim has to be taken with a grain of salt. 
In simple cases  such as the two scalar theories considered in \cite{Held:2025vkd} (see above) 
this equivalence can be established to a certain extent. For QG, however, things are more delicate.
It is known that in this case   Eq.\,\eqref{eq:flow} has a UV-attractive fixed point. Based on their belief   that\,\eqref{pathintegral} and\,\eqref{eq:flow} give equivalent formulations of QG, the authors of \cite{Held:2025vkd} claim that this result  implies the existence of
a specific class $({\rm d}\mu, S)_{\rm grav}$ for
which the path integral\,\eqref{pathintegral} is well-defined in the UV. Actually, they say that $({\rm d}\mu,S)_{\rm grav}$ in\,\eqref{pathintegral} is defined by the fixed point of\,\eqref{eq:flow}, and in section  {\bf A} of \cite{Held:2025vkd}  they  make the bizarre claim  that \vv fixed points define path integral measures''. 

Things work the other way around.  In fact, any QFT (UV-complete or effective) is  defined by $({\rm d}\mu,S)$ through \eqref{pathintegral}. Only when  the theory $({\rm d}\mu,S)$ is  given can the existence of fixed points and their influence on its   UV/IR behavior be investigated.  Let us take the same two theories $({\rm d}\mu_1,S_1)$ and $({\rm d}\mu_2,S_1)$ considered in \cite{Held:2025vkd}. Theory \vv\,1\,'' has two fixed points. One is UV-attractive (Gaussian fixed point), the other  has one relevant and one irrelevant eigendirection (Wilson-Fisher fixed point). Theory \vv\,2\,'' also possesses fixed points, but none of them is UV-attractive. Contrary to what is incorrectly claimed in \cite{Held:2025vkd},  of course this does not mean that theory \vv\,2\,'' is inconsistent. Theory \vv\,2\,'' is a perfectly acceptable EFT.

Let us go back to QG. Taking for $\Gamma_k$ the EH ansatz,  in \cite{Reuter:2001ag} a UV-attractive fixed point of\, \eqref{eq:flow} was found (a similar result was obtained in \,\cite{Bonanno:2004sy} within the proper-time formalism).
Considering the same truncation, 
in \cite{Branchina:2024lai} we reach different conclusions. Our RG equations (42) and (43) for the Newton and cosmological constant $G_k$ and $\Lambda_k$  do not show any sign of such a fixed point.
 
According to the  authors of \cite{Held:2025vkd},
the absence in our RG equations  of this UV-attractive fixed point (AS scenario)
is related to the measure that we use in our work. 
As clearly shown in \cite{Branchina:2024lai}, this in not the point. 
It is true that we take a specific measure\footnote{In \cite{Branchina:2024xzh,Branchina:2024lai} we use the Fradkin-Vilkovisky measure,  proved to be  diffeomorphism invariant in\,\cite{Fradkin:1973wke}. This result was recently questioned in \cite{Bonanno:2025xdg}, where the authors attempt to show that it is rather 
the Fujikawa measure   \cite{Fujikawa:1983im} to be invariant. However, 
non-trivial terms  
are missed in \cite{Bonanno:2025xdg}, and this leads  to not entirely correct results. These issues are discussed in a forthcoming paper \cite{WP1}.}, but the absence of the AS fixed point in our RG equations is due to another reason. To   clarify this point, we recall here the main steps that lead to our RG equations.

As already said, in \cite{Branchina:2024lai} we consider the EH truncation\footnote{Obviously, before moving to higher order terms, the issue of the presence (absence) of the UV-attractive fixed point has to be assessed in this (lowest order) truncation.}	
(as Reuter and collaborators in  \cite{Reuter:2001ag, Bonanno:2004sy}).   
In this framework, we need to compute the quantum corrections to the coefficients of $\sqrt{g}$ and $\sqrt{g} R$ ($g$ is the determinant of the metric, $R$ the Ricci scalar). This calculation is usually performed resorting to the heat-kernel expansion for a generic background metric $\bar g_{\mu\nu}$ \cite{DeWitt:1975ys}. Since we are considering  EH, we can limit ourselves to take as background the metric $g^{(a)}_{\mu\nu}$ of a sphere of radius $a$, as often done in the literature (see for instance \cite{Dou:1997fg}).  In this case, $\int\text{\small${\rm d}^4x\sqrt{g}$}=8\pi^2a^4/3$ \,,\, $\int\text{\small${\rm d}^4x\sqrt{g}$}R=32\pi^2a^2$, and the coefficients of $a^4$ and $a^2$ give the quantum corrections to $\Lambda_{\rm cc}/G$ and $1/G$ respectively.
The RG equation for the running action $S_{\text{\tiny $L$}}[g^{(a)}_{\mu\nu}]$ (Eq.\,(30) of \cite{Branchina:2024lai}) is then derived introducing a (dimensionless) gauge-invariant (running) cut  $L$ on the number of eigenvalues of the Laplacian (see Eqs.\,(20)-(22) of \cite{Branchina:2024lai}). 
In this respect, it is worth to stress that the operators that appear in the fluctuation determinants turn out to be dimensionless (see (19) of \cite{Branchina:2024xzh} and (8) of \cite{Branchina:2024lai}). We will come back to this important point at the end of the present note.

For each  $L$, the dimensionful running scale $k$ is given by (see (25) in \cite{Branchina:2024xzh}, (39) in \cite{Branchina:2024lai}, and comments therein)
\begin{equation}
	\label{correct}
	k = L/a^{\rm dS}_L, 
\end{equation}
where $a^{\rm dS}_L$ is the running de Sitter radius. Let us also observe that, independently of \eqref{correct}, the smallest scale $k$ is the inverse of the universe radius $R_{\text{\tiny U}}$ (see \cite{Polyakov:2000fk}), that means    $k_{\text{\scriptsize min}} \sim 1/R_{\text{\tiny U}} \equiv 1/a^{\rm dS}_{L=1}$. This latter relation is satisfied by \eqref{correct}. 
Should  $a^{\rm dS}_{L}$ in \eqref{correct} be replaced with the off-shell background $a$, we would obtain  $k_{\text{\scriptsize min}}\sim1/a$. However, since $k_{\text{\scriptsize min}}$ is a fixed scale ($k_{\text{\scriptsize min}} \sim R_{\text{\tiny U}}^{-1}$), this dependence on the off-shell radius $a$ appears to be unphysical. 

The RG equations for  $G_k$ and $\Lambda_k$ are finally derived  from the coefficients  of $a^4$ and $a^2$ in the flow equation for $S_L$. Differently from the evolution equations in \cite{Reuter:2001ag,Bonanno:2004sy},  our RG equations (Eqs.\,(42) and (43) of \cite{Branchina:2024lai}) {\it do not} possess the UV-attractive fixed point of the AS scenario.
Contrary to what is claimed in section {\bf A} of \cite{Held:2025vkd}, the absence of this fixed point   is not due to the measure that we use in our work. It is rather due to the relation between $L$ and the running scale  $k$. In fact, if in \eqref{correct} we replace $a^{\rm dS}_L$ with the off-shell background radius $a$ (see however \cite{Branchina:2024xzh,Branchina:2024lai} and comments below \eqref{correct}) while keeping the measure used in our work,  we obtain for $G_k$ and $\Lambda_k$ the 
\vv would-be RG equations'' (65) and (66) of \cite{Branchina:2024lai}. The latter, though derived using the same path integral measure that leads to the RG equations (42) and (43) of \cite{Branchina:2024lai},  
possess the UV-attractive fixed point of the AS scenario (see section 5 of \cite{Branchina:2024lai} for a detailed discussion of this point).

\vskip 3pt

Let us move now to the other incorrect claims of \cite{Held:2025vkd}  (points {\bf B} and {\bf C}), and show that the authors' conclusions are inconsistent. 
They claim (see  section {\bf B} of \cite{Held:2025vkd}) that in our RG equations for  $\Lambda_k$ and $G_k$ (Eqs.\,(42) and\,(43) of \cite{Branchina:2024lai})  we do not account for the implicit scale dependence of the on-shell radius $a_{\rm dS}(k)\equiv a^{\rm dS}_{L}$  (see\,\eqref{correct}). It is immediate to see that this claim is wrong by simply performing the elementary steps (trivial iterated application of the chain rule) that lead from the RG equations (26) and (27)  of \cite{Branchina:2024lai} (written in terms of $L$) to the RG equations (42) and (43) (written in terms of $k$).   This trivial exercise shows that (contrary to what is claimed in \cite{Held:2025vkd}) in \cite{Branchina:2024lai} we consistently derive RG equations that are \vv \,total scale derivatives\,'', and that the zeroes of the corresponding equations for the dimensionless  $\lambda=\Lambda_k/k^2$ and $g=k^2 G_k$ (Eqs.\,(51) and\,(52) of \cite{Branchina:2024lai})  {\it definitely} provide fixed points of the RG flow. 

Another incorrect statement in \cite{Held:2025vkd}  (see sections {\bf B} and {\bf C}) is that we use the background curvature $R=12/a^2$ as \vv \,reference scale\,'' and that we  perform the transition from the metric $g^{(a)}_{\mu\nu}$ on  a sphere of radius $a$ to the metric  $\tilde{g}_{\mu\nu}$ on a unit sphere. They also say that the field redefinition we make 
(see Eq.\,(11) of \cite{Branchina:2024xzh} and also  Eq.\,(6) of \cite{Held:2025vkd})
\begin{equation}\label{redefinition}
	\hat h_{\mu\nu} \equiv (32 \pi G)^{-1/2} \, a^{-1} \, h_{\mu\nu}\,,
\end{equation} 
is the reason why we find a different divergence structure of loop diagrams. 

First of all we stress that in \cite{Branchina:2024lai, Branchina:2024xzh} we never use the background radius $a$ as reference scale. This can be seen for instance in Eq.\,(19) of \cite{Branchina:2024xzh}, where the radius $a$ explicitly appears in the combination $a^2 \Lambda_{\rm cc}$. Moreover,  it is clear 
 that our calculations in \cite{Branchina:2024lai, Branchina:2024xzh}  are performed taking as background the metric  $g^{(a)}_{\mu\nu}$ of a sphere of radius $a$, {\it not} the metric $\tilde g_{\mu\nu}$ of a unit sphere. Actually,  we resort to the identity (in \cite{Branchina:2024lai,Branchina:2024xzh} we use dimensionless coordinates)
\begin{equation}\label{identity}
	g^{(a)}_{\mu\nu}=a^2\tilde g_{\mu\nu}
\end{equation}
simply because it is a convenient (though not necessary) way to dig out the $a$-dependence from the background metric. Needless to say, by no means this implies that  we operate the transition from $g^{(a)}_{\mu\nu}$ to $\tilde g_{\mu\nu}$.  Our results are  obtained with a spherical background of {\it generic radius} $a$. In this respect, we stress that (contrary to what is claimed in \cite{Held:2025vkd}) in our calculation the monomials $R^n$ can certainly be distinguished since they carry different powers of the background radius $a$.

Concerning the field redefinition \eqref{redefinition},  
 the authors of \cite{Held:2025vkd} again miss the point. As for the identity \eqref{identity}, 
it is just a convenient way to collect the $a$-dependence. Obviously, it is not a necessary step of the calculation, and the same results are obtained even if this  redefinition is not introduced. Therefore, contrary to what is claimed in section {\bf C} of \cite{Held:2025vkd}, the different (compared to usual results) divergence structure that we find in \cite{Branchina:2024xzh} is not due to\,\eqref{redefinition}. It  comes from the fact that we pay due attention to the path integral measure and to the identification of the physical UV cutoff.

Before ending this note, we would like to point out another important result of \cite{Branchina:2024lai, Branchina:2024xzh}.  There  we show that the quantum fluctuation operators turn out to be {\it  dimensionless}. The authors of \cite{Held:2025vkd} make an incorrect claim also on this point.  They say that  the dimensionless nature of these operators is due to the fact that we \vv make a transition'' from the metric $g^{(a)}_{\mu\nu}$  of a sphere of radius $a$ to the metric  $\tilde g_{\mu\nu}$ of a unit sphere (see their section {\bf C}). As stressed above, the appearance of $\tilde g_{\mu\nu}$ has nothing to do with such an alleged transition.
We repeat here the point. It appears in our expressions simply because we  write  $g^{(a)}_{\mu\nu}$ in a way that is convenient (but not necessary) for the calculation. A simple reading of our works \cite{Branchina:2024lai, Branchina:2024xzh} shows  that our calculations are performed for a sphere of generic radius $a$.
\vskip 3pt

{\bf Summary} -  In the present note we have shown that the arguments put forward by the authors of \cite{Held:2025vkd}
against our work \cite{Branchina:2024xzh,Branchina:2024lai}
are inconsistent and their claims incorrect.  In particular, we have shown that contrary to what the authors of \cite{Held:2025vkd} claim: (i) the absence of the UV-attractive fixed point of the asymptotic safety scenario in our RG equations, Eqs.\,(42) and (43) of \cite{Branchina:2024lai}, is not due to the measure we use; (ii) our beta functions, Eqs.\,(51) and (52) of \cite{Branchina:2024lai}, are {\it total}
scale derivatives and their zeroes {\it do define} fixed
points; (iii) the calculations in  \cite{Branchina:2024xzh, Branchina:2024lai} are performed for a sphere of generic radius $a$, not for a unit sphere. Moreover, we have shown that 
their claim \vv fixed points define path integral measures'' is meaningless: according
to Wilson's paradigm, the existence of fixed points can be assessed only when the theory $({\rm d}\mu,S)$ is  given.

\subsection*{Acknowledgments}
 
The work of CB has been supported
by the European Union – Next Generation EU through the research grant number P2022Z4P4B
“SOPHYA - Sustainable Optimised PHYsics Algorithms: fundamental physics to build an advanced
society” under the program PRIN 2022 PNRR of the Italian Ministero dell’Università e Ricerca
(MUR). The work of VB, FC, RG and AP is carried out within the INFN project QGSKY.

\begin{acronym}[EFT]
\acro{UV}[UV]{ultraviolet}
\acro{QFT}[QFT]{quantum field theory}
\acrodefplural{QFT}{quantum field theories}
\acro{RG}[RG]{renormalization group}
\end{acronym}

\end{document}